# Discovery of a New Molecular Bubble–Outflow Structure in the Taurus B18 Cloud

Yan Duan[1,2], Di Li[1,3,4], Paul F. Goldsmith[5], Laurent Pagani[6], Tao-Chung Ching[1,7], Shu Liu[1], Jinjin Xie[8], and Chen Wang[1]
[1] National Astronomical Observatories, Chinese Academy of Sciences, Beijing 100101, People's Republic of China; dili@nao.cas.cn
[2] University of Chinese Academy of Sciences, Beijing 100049, People's Republic of China
[3] Research Center for Intelligent Computing Platforms, Zhejiang Laboratory, Hangzhou 311100, People's Republic of China
[4] NAOC–UKZN Computational Astrophysics Centre, University of KwaZulu-Natal, Durban 4000, South Africa
[5] Jet Propulsion Laboratory, California Institute of Technology, 4800 Oak Grove Drive, Pasadena, CA 91109, USA
[6] LERMA & UMR 8112 du CNRS, Observatoire de Paris, PSL University, Sorbonne Universités, CNRS, F-75014 Paris, France
[7] Zhejiang Lab, Hangzhou, Zhejiang 311121, People's Republic of China
[8] Shanghai Astronomical Observatory, Chinese Academy of Sciences, 80 Nandan Road, Shanghai 200030, People's Republic of China
*Received 2022 June 28; revised 2022 November 30; accepted 2022 November 30; published 2023 February 7*

## Abstract

Star formation can produce bubbles and outflows, as a result of stellar feedback. Outflows and bubbles inject momentum and energy into the surrounding interstellar medium, and so are related to the overall energy balance of the molecular cloud. Molecular bubbles can be resolved by higher-resolution radio telescopes to quantify the effect of star formation on molecular clouds. We report here the identification of a new molecular bubble with an outflow, and a Herbig–Haro object, HH 319, located at the bubble center. Multiwavelength data have been utilized to study its spatial structure, energy injection, and dynamical timescale. This bubble has a kinetic energy of $5.8 \times 10^{43}$ erg within the smallest radius of a bubble in Taurus, 0.077 pc. The bubble formed ∼70,000 yr ago. According to the proper-motion velocities of protostars from Gaia EDR3, the T Tauri binary stars (FY Tau and FZ Tau) at the southwest edge of the bubble may have produced the outflow–bubble structure. This is an unusual new structure found in low- and intermediate-mass star formation regions. Only a bubble in Orion A, driven by V380 Ori, has a similar structure. The bubble–outflow structure provides additional observational evidence for the theory of stellar wind from T Tauri stars. It enhances our understanding of how stellar feedback acts on molecular clouds.

*Unified Astronomy Thesaurus concepts:* Stellar wind bubbles (1635); Stellar winds (1636); Molecular clouds (1072); T Tauri stars (1681); Stellar feedback (1602)

## 1. Introduction

Star formation is related to the structure and evolution of galaxies and giant molecular clouds, the synthesis of chemical elements, and the star formation rate and is intimately connected with a critical astrophysical parameter, the initial mass function (McKee & Ostriker 2007). Almost all young stellar objects (YSOs) have undergone periods of substantial mass loss. YSOs' fast and collimated stellar wind sweeps through the surrounding molecular gas. The molecular gas is expelled from the cavities and expands as irregular lobes and incomplete shells (Bachiller 1996). These become visible as molecular bubbles and molecular outflows. Herbig–Haro (HH) objects, consisting of knots of ionized gas, are the optical manifestations of this mass-loss process (Reipurth & Bally 2001; McKee & Ostriker 2007).

As a very common phenomenon in both high-mass and low-mass star formation regions, molecular outflows and molecular bubbles are two forms of stellar feedback (e.g., Lada 1985; Arce et al. 2007; Li et al. 2015). Both outflows and bubbles are important in the evolution of the surrounding interstellar medium, injecting momentum and energy into the cloud. Even very low-mass objects can inject energy into the molecular cloud (Phan-Bao et al. 2008). In the Taurus molecular cloud complex, the energy injection rate from the bubbles and outflows can match the turbulence dissipation rate (Li et al. 2015). Molecular bubbles have a greater impact on molecular clouds than outflows. The dynamical timescales of bubbles are longer than those of outflows in Taurus (Li et al. 2015). Thus, the energy injection from bubbles into the molecular clouds could last longer. The ensemble of bubbles in Taurus have about 110 times greater mass and 24 times higher energy than the total of the outflows (Li et al. 2015). Compared to supernova remnants, molecular bubbles could exist for a longer period of time (Arce et al. 2011). Due to their complex morphologies, bubbles are more difficult to identify than relatively clear outflow features (Arce et al. 2011). Therefore, the study of molecular bubbles is both important and challenging.

Bubbles arise from a largely spherical stellar wind. Observationally, bubbles are similar to complete or partial rings. Arce et al. (2011), Li et al. (2015), and Feddersen et al. (2018) have completed a systematic effort to identify bubbles in the Perseus molecular cloud complex, Taurus molecular cloud complex, and Orion A giant molecular cloud. From low- and intermediate-mass star formation regions to massive star formation regions, the minimum mass of a detectable bubble is $2\,M_{\odot}$ with an energy of $2 \times 10^{43}$ erg. The minimum radii of the bubbles are 0.28 pc and 0.050 pc for Taurus and Orion, respectively (Li et al. 2015; Feddersen et al. 2018). The previous bubble identifications in Taurus and Perseus are energetically complete (Liu et al. 2019). Current observations of the bubbles on small spatial scales are still limited by the telescopes' sensitivities and sky coverages. Bubbles from low-mass stars are capable of being resolved by higher-resolution telescopes. Studies of bubbles help us understand the variability of energy injection and spatial structure, the evolution of YSOs, and the evolution of stellar wind.

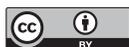

Protostellar outflows are traced by radiative shock waves at near-UV, visible, and infrared wavelengths, or by high-velocity line wings on submillimeter to millimeter wavelength emission lines (Bally 2011). Extensive outflow surveys have been conducted in the Perseus, Ophiuchus, and Taurus molecular clouds (Arce et al. 2010; Nakamura et al. 2011a, 2011b; Li et al. 2015). In the Taurus molecular cloud complex, the minimum mass of outflows is 0.001 $M_\odot$ and the minimum energy is $1 \times 10^{41}$ erg (Li et al. 2015).

Increasing evidence suggests that the collimation and morphology of the outflow change with time (e.g., Arce & Sargent 2006; Seale & Looney 2008; Velusamy et al. 2014). At an early evolutionary stage, a young protostar could launch a collimated wind and jetlike outflow (e.g., Arce et al. 2007; Frank et al. 2014; Bally 2016). More evolved pre-main-sequence stars drive wide-angle or spherical winds (e.g., Castor et al. 1975; Bally 2011). Pre-main-sequence stars, such as T Tauri stars and Herbig Ae/Be stars, drive wide-angle winds that coexist with a collimated jetlike wind component (Arce et al. 2011).

At a distance of 140 pc (Torres et al. 2009), the Taurus molecular cloud complex is a widely studied low-mass star-forming region. The B18 cloud (also known as TMC-2; Onishi et al. 1996) is located in the southern part of Taurus. The central part of B18 contains dozens of young stars and several HH objects. We have detected a new molecular bubble in the Taurus B18 cloud. A previously identified outflow is located at the center of the bubble (Li et al. 2015). The present observations allow us to derive the properties and potential interplay of the bubble and the outflow. With our analysis of the structure, estimation of the energy injection, and analysis of the formation history, we can understand the remarkable effects of low-mass star formation on a surrounding cloud.

In Section 2, we present the basic information about the IRAM 30 m observations. In Section 3, we discuss the spatial structure, velocity information, previous studies, and physical parameter estimation of the bubble, the outflow, and the HH objects. In Section 4.1, we discuss the possible driving sources and formation history of this structure. In Section 4.2, we compare our source with existing observations and models. In Section 5, we summarize the main conclusions of this paper.

## 2. Observations

We performed $^{12}$CO, $^{13}$CO, and C$^{18}$O $J = 2 \rightarrow 1$ observations toward the bubble by the IRAM 30 m telescope and $^{12}$CO $J = 3 \rightarrow 2$ observations toward the Taurus B18 cloud with the James Clerk Maxwell Telescope (JCMT). We collected multiwavelength data and summarize the images of the bubble–outflow structure in radio, infrared, and optical wavelengths and show the position of this structure in the Taurus molecular cloud in Figure 1. We present the IRAM 30 m $^{12}$CO, $^{13}$CO, and C$^{18}$O $J = 2 \rightarrow 1$ observations in Section 2.1 and the JCMT $^{12}$CO $J = 3 \rightarrow 2$ observations (Y. Duan et al. 2023, in preparation) in Section 2.2.

### 2.1. $^{12}$CO J = 2 → 1 from IRAM 30 m Telescope

The observations were performed with the IRAM 30 m telescope on January 10, 11, and 12 in 2019 (program ID: 161-18). We observed the $^{12}$CO, $^{13}$CO, and C$^{18}$O $J = 2 \rightarrow 1$ emission in a $6' \times 5'$ region centered at $4^h32^m35^s+24°21'00''$ (J2000). The rest-frame frequencies are 230.538000 GHz, 220.398684 GHz, and 219.560358 GHz for $^{12}$CO, $^{13}$CO, and C$^{18}$O $J = 2 \rightarrow 1$, respectively. The half-power beamwidths (HPBWs) at the three frequencies are $10\rlap{.}''7$, $11\rlap{.}''2$, and $11\rlap{.}''2$, respectively. The sensitivities of the antenna temperature for $^{12}$CO, $^{13}$CO, and C$^{18}$O $J = 2 \rightarrow 1$ are 0.35 K, 0.21 K, and 0.21 K for a velocity resolution of 0.25 km s$^{-1}$, 0.27 km s$^{-1}$, and 0.27 km s$^{-1}$, respectively. The spectra were treated and baseline-fitted using the GILDAS software.[9] The main-beam efficiency is 0.59 at 230 GHz. The selected OFF position for these observations was set incorrectly and shows some emission. As a consequence, the flux of the $^{12}$CO and $^{13}$CO observations is partly missing. We observed the OFF position independently to correct the data. We refer the reader to Appendix A for details. The shown data include the correction.

### 2.2. $^{12}$CO J = 3 → 2 from JCMT

We utilized a 1.3 deg$^2$ $^{12}$CO $J = 3 \rightarrow 2$ mapping of the Taurus B18 cloud from a 14 hr JCMT–HARP observation (Y. Duan et al. 2023, in preparation). The observations were performed in band 3 on 2017 September 6, 11, and 13; 2017 November 14; and 2018 August 10 (program ID: M17BP027 and M18BP072). $^{12}$CO $J = 3 \rightarrow 2$ has a rest frequency of 345.79599 GHz and the selected spectral resolution is 61 kHz (0.05 km s$^{-1}$). The telescope's HPBW is 14″ at this frequency. The sensitivity of the antenna temperature is 1.5 K for a velocity resolution of 0.05 km s$^{-1}$. The data were processed with the Starlink package (Currie et al. 2014).

## 3. Result and Analysis

In the $^{12}$CO $J = 3 \rightarrow 2$ map of Taurus B18 observed by JCMT (Y. Duan et al. 2023, in preparation), we discovered a new molecular bubble and acquired IRAM $^{12}$CO, $^{13}$CO, and C$^{18}$O $J = 2 \rightarrow 1$ mapping. Combining this mapping with the multiwavelength data, we conduct analysis of the bubble (in Section 3.1), the central outflow (in Section 3.2), the HH object (Haro 6–19 in Haro 1953, now known as HH 319; in Section 3.3), and the physical parameter estimation (in Section 3.4).

A three-color map of the identified bubble is shown in Figure 2, exhibiting the structure of the bubble, outflow, and YSOs at optical, infrared, and radio wavelengths. The three colors represent the integrated intensity maps of the $^{13}$CO (red), 250 $\mu$m (green), and H$\alpha$ (blue) images, which are in panels (j), (g), and (c) of Figure 1, respectively. The H$\alpha$ image was obtained from the NOIRLab Astro Data Archive (formerly the NOAO Science Archive).[10] In Figure 1(c), there is H$\alpha$ emission tracing HH 319, marked as a white cross. At the southwest of the bubble, there is strong H$\alpha$ emission near the T Tauri binary stars (FY Tau and FZ Tau), marked as two white stars. In Figure 2, the distributions of the molecular gas traced by the $^{13}$CO 5–6 km s$^{-1}$ integrated intensity map and the dust traced by the Herschel 250 $\mu$m image are not coexistent. This may be due to the $^{13}$CO and C$^{18}$O images in velocities greater and less than 6 km s$^{-1}$ showing an east–west symmetrical distribution (see Figure 4 for details). The $^{13}$CO and C$^{18}$O data integrated from 6 to 7 km s$^{-1}$ should be associated with dust.

---

[9] https://www.iram.fr/IRAMFR/GILDAS/
[10] https://astroarchive.noirlab.edu

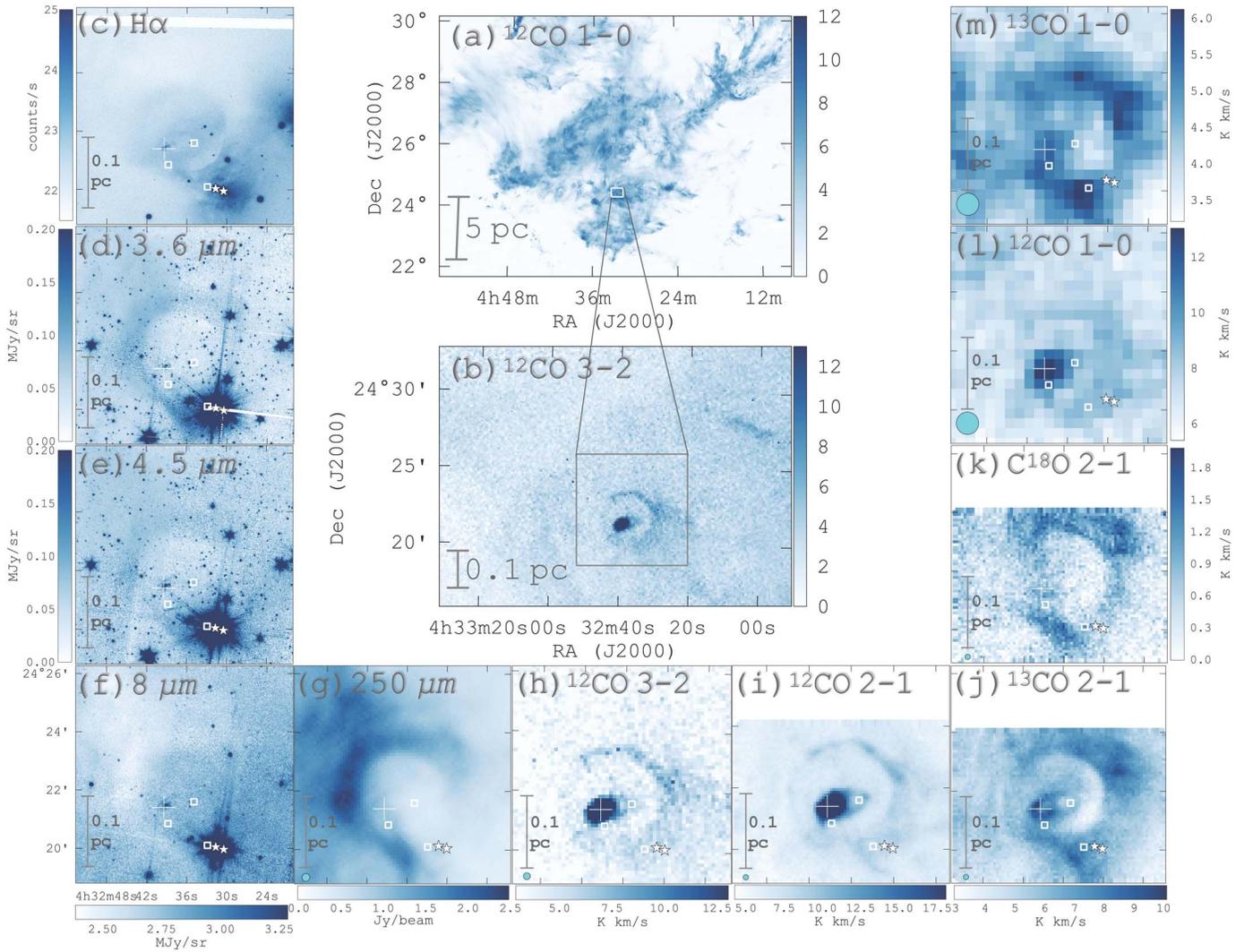

**Figure 1.** (a) Integrated intensity map of the Taurus molecular cloud, as traced by $^{12}$CO $J = 1 \to 0$ from the FCRAO 13.7 m telescope (Goldsmith et al. 2008; Narayanan et al. 2008). (b) Integrated intensity map of B18 cloud in Taurus, as traced by $^{12}$CO $J = 3 \to 2$ from JCMT (unpublished data; Y. Duan et al. 2023, in preparation). (c)–(m) Zoomed-in bubble–outflow images from the gray box region shown in panel (b). The sky coverages in panels (c)–(m) are the same. The images are an H$\alpha$ image (c) from the NOIRLab Astro Data Archive; Spitzer 3.6 $\mu$m (d), 4.5 $\mu$m (e), and 8 $\mu$m (f) images (Fazio et al. 2004; Werner et al. 2004; IRSA 2022); a Herschel 250 $\mu$m image (g) (Eales et al. 2010; Valiante et al. 2016; H-ATLAS Team 2020); a $^{12}$CO $J = 3 \to 2$ image from JCMT (h) (Y. Duan et al. 2023, in preparation); $^{12}$CO (i), $^{13}$CO (j), and C$^{18}$O $J = 2 \to 1$ (k) images from the IRAM 30 m telescope; and $^{12}$CO (l) and $^{13}$CO $J = 1 \to 0$ (m) images from the $\sim$100 deg$^2$ FCRAO CO survey in the Taurus molecular cloud (Goldsmith et al. 2008; Narayanan et al. 2008). All the CO maps are integrated in the velocity range of 5–6 km s$^{-1}$. The blue circles in the lower left corner of panels (g)–(m) are the beam sizes. HH 319 has a knotty structure. The white cross and squares are HH 319 and other HH objects presented by Magakian et al. (2002) (see Section 3.3 for details). The white stars are the T Tauri binary stars (FY Tau and FZ Tau).

For the 5 to 6 km s$^{-1}$ plots, the molecular gas is symmetrically distributed with respect to the dust emission.

### 3.1. Molecular Bubble

For the bubble survey of molecular clouds, we followed previous studies using similar steps to do the identification (Arce et al. 2010; Li et al. 2015; Feddersen et al. 2018). The steps in the commonly employed empirical procedure used to identify bubbles are summarized by Liu et al. (2019) as follows: (1) Search for bubble-like structures brighter than the surrounding molecular gas. (2) Draw the position–velocity diagram (hereafter P–V diagram) and search for "V" or "∪" shapes in the figure. An expanding bubble should appear with redshifted or blueshifted gas shown in the P–V diagram. (3) Search for nearby protostars as the driving source. (4) Search the infrared data to see if the bubble shapes are similar to CO data in the corresponding positions.

Here, we can see the bubble-like structure in the $^{12}$CO, $^{13}$CO, and C$^{18}$O integrated intensity maps. Due to the influence of the outflow, the P–V diagram is more complicated. We draw the P–V diagram in Figure 3 avoiding the location of the outflow. We cannot distinguish clear "V" or "∪" shapes in the P–V diagram. The $^{13}$CO gas in the bubble interior is redshifted and the bubble is blueshifted. This slight trend tends to form a "V" shape in the P–V diagram. We searched for nearby protostars and discuss them in Section 4.1.2. We checked the Herschel SPIRE map at 250 $\mu$m and found that the bubble is traced by the dust continuum (see Figures 1(d) and 2). The bubble formation pushed the dust away and formed a cavity. Using the above four steps, we identified this molecular bubble.

The channel maps of the $^{12}$CO, $^{13}$CO, and C$^{18}$O $J = 2 \to 1$ data are shown in Figure 4 for the bubble and in Figure 5 for the outflow. The high-resolution mapping reveals a clear structure of the outflow located at the center of the bubble. In the $^{12}$CO channel maps, the bubble presents a clear ring shape. To maximize the contrast, we drew the $^{13}$CO and C$^{18}$O maps in the range of 5–6 km s$^{-1}$, showing the asymmetrical distribution of the surrounding cloud.

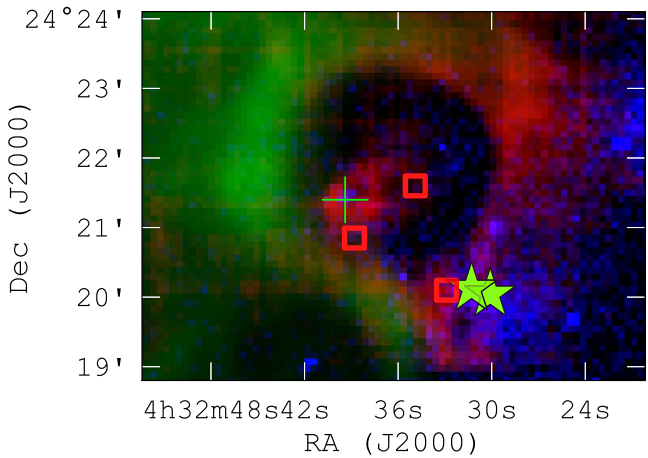

**Figure 2.** Three-color map of the bubble–outflow structure. The red represents the $^{13}$CO integrated intensity. The velocity range is 5–6 km s$^{-1}$ (range selected to provide the best contrast). The green represents the Herschel 250 $\mu$m image (Eales et al. 2010; Valiante et al. 2016; H-ATLAS Team 2020). The blue represents the H$\alpha$ image. We mark HH 319 and other HH objects as a green cross and red squares, respectively. The green stars represent the pair of T Tauri stars.

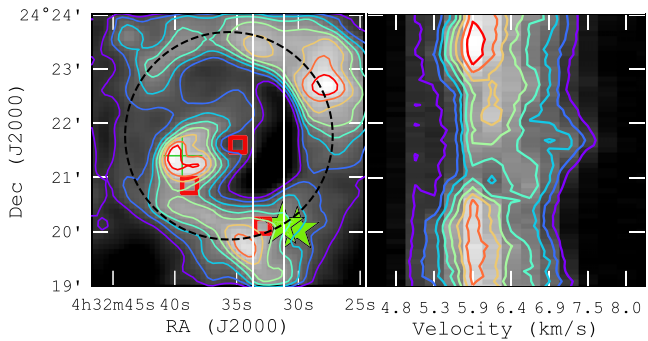

**Figure 3.** Left: gray-scale and contour map showing the integrated intensity map of $^{13}$CO $J = 2 \rightarrow 1$. The integrated velocity range is 5–6 km s$^{-1}$. The contour levels are 60%, 65%, ..., 95% of the peak value. The data are smoothed to 20″. The solid white lines represent the cut for the P–V diagram shown in the right panel. The black dashed line is the circle from the fitted radius (discussed in Section 3.4). The green cross and red squares are HH 319 and other HH objects. The green stars are the T Tauri binary stars (FY Tau and FZ Tau). Right: the $^{13}$CO $J = 2 \rightarrow 1$ P–V diagram along the decl. direction of a 36″ wide slice. This interval was chosen to avoid the area of the outflow. The contour levels are 20%, 30%, ..., 90% of the peak value.

We analyzed the Spitzer images and discovered that the bubble is brighter than its surroundings at 3.6 $\mu$m and 4.5 $\mu$m and can be seen in absorption at 8 $\mu$m in panels (d)–(f) of Figure 1. We show a three-color map of the three bands in Figure 6. This could be a case of "coreshine" emission (Pagani et al. 2010), caused by the scattering of infrared starlight from relatively large (~0.5–1 $\mu$m radius) grains. The existence of the "coreshine" effect signifies an inner dense region lacking any UV or optical illumination.

The east side of this bubble, which is the original unperturbed material, is traced by the Herschel and Spitzer images. The cloud has been locally destroyed by stellar feedback, creating a rupture in the west, which is also seen as a clear deficit in the molecular gas. The stellar wind sweeps up the gas and creates a cavity at multiple wavelengths.

### 3.2. The Central Outflow

The outflow has been studied before. Lichten (1982) first reported the detection of a typically broad, low-intensity $^{12}$CO $J = 1 \rightarrow 0$ line wing there. The line wing at $4^h32^m45^s$ +24°25′ 18″ (J2000) has a full width of 30 km s$^{-1}$ (Lichten 1982). The rms noise level is 0.09 K. The collimated bipolar outflow is considered an explanation for the broad molecular line wings. There is no specific anisotropic spatial distribution of red and blue lobes of an outflow. Lichten (1982) did not confirm the presence of a bipolar outflow. Using $^{12}$CO $J = 1 \rightarrow 0$ mapping with a higher resolution than that of Lichten (1982), Li et al. (2015) first identified the existence of a bipolar outflow here, named TMO_16 (SST 043231.7 + 242002). We detected this bipolar outflow with IRAM $^{12}$CO $J = 2 \rightarrow 1$ data.

We can see the high-speed gas flow in the P–V diagram of the outflow in Figure 7. In the P–V diagram, the blue lobe shows a high velocity of gas up to 1.9 km s$^{-1}$, at 24°21′–24° 22′. The red lobe is weaker than the blue lobe. At 24°22′, the red lobe shows a faint gas of high velocity, greater than 6.5 km s$^{-1}$. The spatial distribution of the red lobe is more extended and weaker than that of the blue lobe. The average spectra in Figure 7 of the two lobes reveal the broad line wings of the outflow.

### 3.3. HH 319 in the Outflow

HH 319 is located at the center of the outflow, providing crucial information about its origin. In the H$\alpha$ and [S II] images, HH 319 has an elongated, arrowhead-like shape and consists of several knots. Magakian et al. (2002) identified seven knots of HH 319 (named "HH 319A"–"HH 319G") and three adjacent independent HH objects (named "Obj.1," "Obj.2," and "Obj.3"; see their Figures 1 and 2). The coordinates of all the HH objects were given (see Magakian et al. 2002, Table 1). In the entire text, HH 319 is represented as a cross. The three other HH objects are represented as squares.

A weak bubble in the H$\alpha$ image (see Figure 1 in Magakian et al. 2002) surrounds HH 319, at the same position of the molecular bubble, which is probably due to the shock. Magakian et al. (2002) considered the T Tauri stars (FY Tau and FZ Tau) could be the possible driving sources of HH 319. The position angle of the proper motion of HH 319 is 72°. The angle from the plane of the sky is 53° (Magakian et al. 2002). The proper-motion velocity of HH 319 is 0″.05/yr (Luyten 1971), which corresponds to a tangential velocity of 30 km s$^{-1}$ at the distance of 140 pc of Taurus (Magakian et al. 2002).

Knot A in HH 319 (refer to Figure 2 in Magakian et al. 2002) is the brightest part of HH 319, located at the apex of the shock. Throughout the text, the position of HH 319A represents HH 319. HH 319A is located in the southeast direction. To comply with its direction, the driving source of HH 319 must be located in the northwest, inconsistent with the position of the T Tauri stars. We explain that by giving a possible proper-motion trajectory of the T Tauri stars in Section 4.1.2.

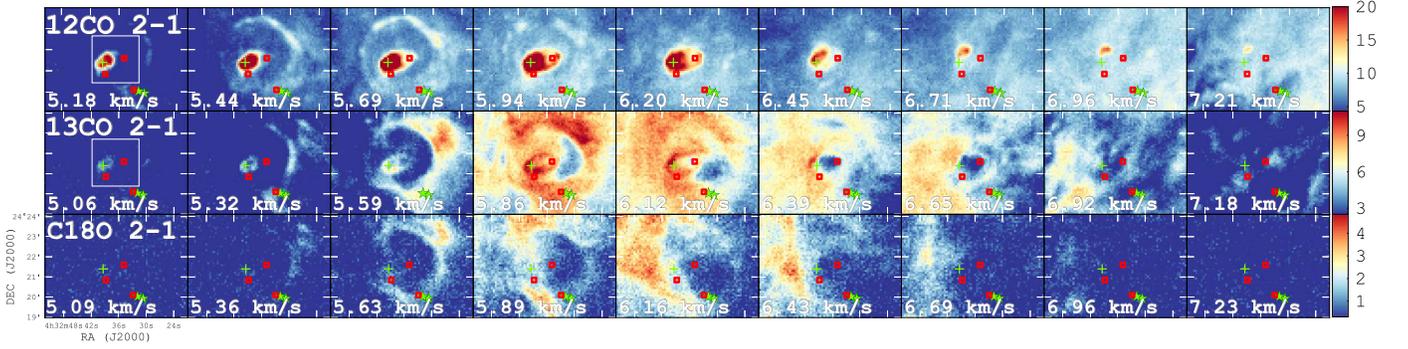

**Figure 4.** Channel maps of $^{12}$CO, $^{13}$CO, and C$^{18}$O $J = 2 \to 1$ from IRAM observations. The central velocity of each channel is indicated in the lower left corner. The white box in the leftmost $^{12}$CO and $^{13}$CO images delineates the zoomed-in images displayed in Figure 5. The green cross and red squares are HH 319 and other HH objects presented by Magakian et al. (2002). The green stars are the T Tauri binary stars.

### 3.4. Physical Parameters

To compare the physical parameters of the bubble with the total Taurus bubble inventory, we follow the method of Li et al. (2015). The physical parameters of the outflow and the bubble are summarized in Table 1.

In Table 1, $\Delta V$ has different significations for the outflow and the bubble. The outflow velocity, $\Delta V = |V_{obs} - V_{sys}|$, is the average observed velocity of the outflow ($V_{obs}$) relative to the cloud systemic velocity ($V_{sys}$) (Arce et al. 2010; Li et al. 2015). $V_{sys}$ is from the Gaussian fitting of $^{13}$CO, which is 6.2 km s$^{-1}$. The observed average velocities of the blue and red lobes are 3.6 km s$^{-1}$ and 8.4 km s$^{-1}$, respectively. For the outflow, $\Delta V$(blue) = 2.6 km s$^{-1}$ and $\Delta V$(red) = 2.2 km s$^{-1}$. For the molecular bubble, $\Delta V$ is the expansion velocity of the bubble determined by visual inspection in channel maps (Li et al. 2015).

$S$ in Table 1 is the area of the bubble and of the outflow projected onto the sky plane. For the outflow, the blue lobe is better resolved with the resolution of 11″ of the IRAM 30 m telescope. We estimate the area of the blue lobe as $0\rlap{.}'8 \times 0\rlap{.}'8$ rather than $2' \times 2'$ from Li et al. (2015). For the red lobe of the outflow, the gas structure is extended. We confirm the $4' \times 4'$ size found by Li et al. (2015).

The distance of Taurus is 140 pc (Torres et al. 2009). We measure the maximum length of the outflow, denoted as $\lambda_{max}$. $R_{bubble}$ is the radius of the bubble, which is fitted by least-squares fitting. The $^{12}$CO $J = 2 \to 1$ image in the 5.69 km s$^{-1}$ velocity channel is cut into a $5 \times 5$ array of equal-sized tiles. The brightest point of each tile is selected and points close to the bubble periphery are manually selected and used to fit the circle. The fitted center of the circle is $4^h32^m35\rlap{.}^s6 +24°21'46''$ (J2000) with a radius of 0.077 pc.

We use the $^{13}$CO and $^{12}$CO data to calculate the column density of the bubble and the outflow, respectively. For the outflow, we use $^{12}$CO $J = 2 \to 1$ data integrated over 2.0–4.5 km s$^{-1}$ for the blue lobe and 7.5–9.0 km s$^{-1}$ for the red lobe. For the bubble, we use a $^{13}$CO $J = 2 \to 1$ map integrated over 4.7–8.0 km s$^{-1}$. The detailed derivations of column density ($N_{tot}$), mass ($M_{gas}$), momentum ($P$), energy ($E$), and energy injection rate ($\dot{E}$) for the outflow and the bubble are given in Appendix B. Referring to the Taurus outflow and bubble survey (Li et al. 2015), the excitation temperature, $T_{ex}$, is assumed to be 25 K. We assume that $^{12}$CO is optically thick and $^{13}$CO is optically thin in the calculations of the column densities. We summarize the results in Table 1.

Among the 37 bubbles in Taurus studied by Li et al. (2015), the minimum and median radii are 0.28 pc and 0.70 pc, respectively. The IRAM 30 m telescope at CO $J = 2 \to 1$ has one-quarter the beam size of the FCRAO 13.7 m telescope at CO $J = 1 \to 0$. Our CO $J = 2 \to 1$ molecular bubble has a radius of 0.077 pc, which is smaller than any in the Taurus bubble survey by FCRAO. The minimum and median masses of the 37 bubbles in Taurus studied by Li et al. (2015) are 2 $M_\odot$ and 12 $M_\odot$, respectively. The minimum and median momenta are 2 $M_\odot$ km s$^{-1}$ and 21 $M_\odot$ km s$^{-1}$, respectively. The minimum and median energy are $2 \times 10^{43}$ erg and $41 \times 10^{43}$ erg, respectively. The minimum and median energy injection rates are $0.20 \times 10^{31}$ erg s$^{-1}$ and $3.1 \times 10^{31}$ erg s$^{-1}$, respectively. Our molecular bubble has a mass and momentum of 5.2 $M_\odot$ and 5.5 $M_\odot$ km s$^{-1}$, respectively. The bubble injects $5.8 \times 10^{43}$ erg and the energy injection rate is $2.6 \times 10^{31}$ erg s$^{-1}$. The mass, momentum, energy, and energy injection rate of this molecular bubble are all lower than the median value of the bubble survey and slightly higher than the lowest value of the survey.

The physical parameters of this outflow (named "TMO_16") have been published by Li et al. (2015). With a higher-resolution mapping, we obtain a new estimate of the physical parameters for the outflow. The 56 outflows in Taurus identified by Li et al. (2015) have minimum and median lengths of 0.1 pc and 0.3 pc, respectively. The maximum length of our outflow is 0.046 pc and 0.23 pc for the blue lobe and red lobe, respectively. The red lobe is close to the median length, and the blue lobe is smaller than the minimum value. It is reasonable that a higher resolution could reveal the subtle structure of the same source. The minimum mass and median mass for the 55 outflows in Taurus are 0.001 $M_\odot$ and 0.034 $M_\odot$, respectively. The minimum and median momenta are 0.003 $M_\odot$ km s$^{-1}$ and 0.098 $M_\odot$ km s$^{-1}$, respectively. The minimum and median energies are $0.01 \times 10^{43}$ erg and $0.26 \times 10^{43}$ erg, respectively. The minimum and median energy injection rates are $0.004 \times 10^{31}$ erg s$^{-1}$ and $0.053 \times 10^{31}$ erg s$^{-1}$, respectively (Li et al. 2015). The mass, momentum, and energy of the red lobe for this outflow are close to the median values of the outflow survey, and those of the blue lobe are slightly higher than the lowest values of the outflow survey. The energy injection rates of the blue lobe and red lobe are close to the median level.

To estimate the dynamical timescale, we expect that the protostellar outflow and bubble formed in a relatively quiescent cloud, and that they probably formed at different times. The

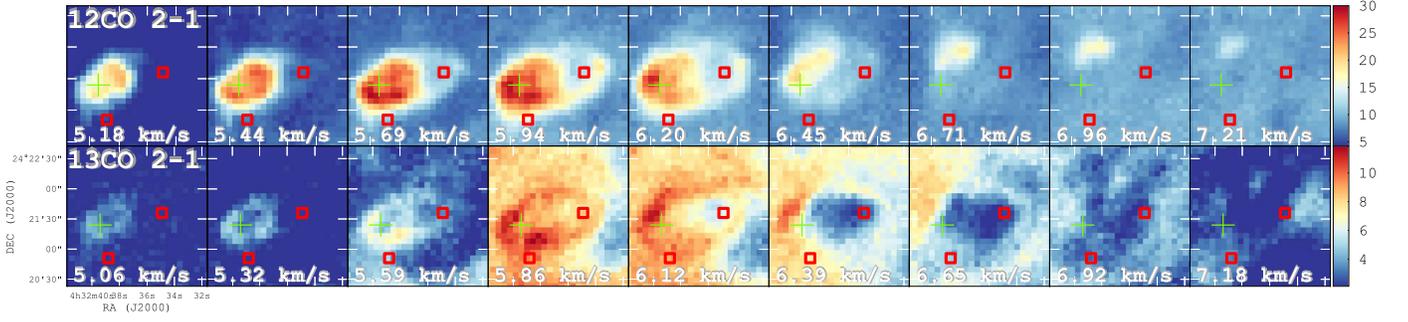

**Figure 5.** Zoomed-in $^{12}$CO and $^{13}$CO images from the white box region shown in Figure 4. The markers are the same as those in Figure 4.

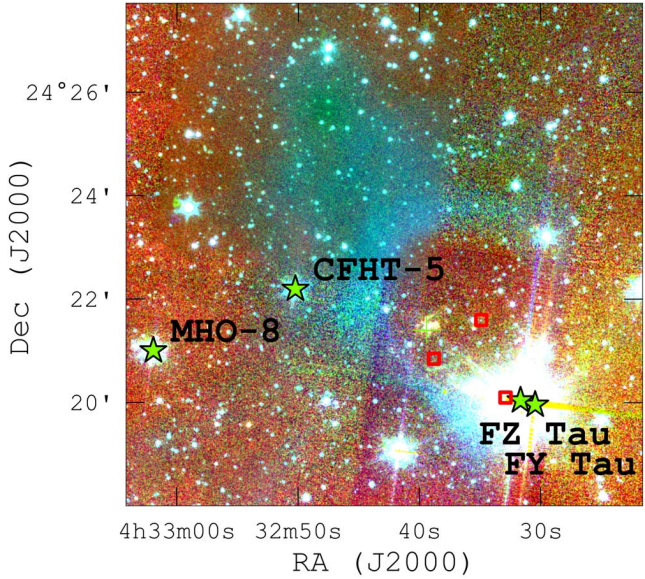

**Figure 6.** Three-color map of Spitzer 3.6 μm (blue), 4.5 μm (green), and 8 μm (red) IRAC channels (Fazio et al. 2004; Werner et al. 2004; IRSA 2022). We show the 3.6 μm, 4.5 μm, and 8 μm maps of the bubble in panels (d), (e), and (f) of Figure 1, respectively. The markers are the same as those in Figure 4.

dynamical timescale ($t_{dyn}$) can be estimated by

$$t_{dyn(outflow)} = \frac{\lambda_{max}}{\Delta V}, \quad t_{dyn(bubble)} = \frac{R_{bubble}}{\Delta V}. \quad (1)$$

In Table 1, we give the estimated $t_{dyn}$ for the bubble and the two lobes of the outflow.

Considering the inclination of the outflow axis to the line of sight, the dynamical timescales corrected for the projection angle ($t_{dyn,corr}$) are

$$t_{dyn,corr(outflow)} = \frac{\lambda_{max}/\sin(i)}{\Delta V/\cos(i)} = t_{dyn}/\tan(i), \quad (2)$$

where $i$ is the angle to the line of sight. Similarly, we calculate the corrected values of the momentum ($P_{corr}$), energy ($E_{corr}$), and energy injection rate ($\dot{E}_{corr}$) of the outflow, through the relations

$$P_{corr(outflow)} = M_{gas}\Delta V_{corr} = M_{gas}\Delta V/\cos(i), \quad (3)$$

$$E_{corr(outflow)} = \frac{1}{2}M_{gas}\Delta V_{corr}^2 = \frac{1}{2}M_{gas}\Delta V^2/\cos^2(i), \quad (4)$$

$$\dot{E}_{corr(outflow)} = \frac{E_{corr}}{t_{dyn,corr}}. \quad (5)$$

In Magakian et al. (2002), the angle of the HH 319 flow with respect to the plane of the sky is 53°. The angle between HH 319 and the line of sight should be 37°. We assume that the inclination angle of the outflow with respect to the line-of-sight direction coincides with the angle of HH 319. On the basis of the inclination angle, 37°, the $P_{corr}$, $E_{corr}$, $t_{dyn,corr}$, and $\dot{E}_{corr}$ of the outflow are corrected to 1.3 times, 1.6 times, 1.3 times, and 1.2 times the uncorrected values, respectively. A molecular bubble should be a 3D sphere in space and its dynamical timescale requires no correction. We do not consider the effect of the sky projection for the bubble here. In Table 1, only the momentum, energy, dynamical timescale, and energy injection rate of the outflow ($P_{corr}$, $E_{corr}$, $t_{dyn,corr}$, and $\dot{E}_{corr}$) are modified by the 37° inclination angle of the line-of-sight direction. The rest of the physical parameters in Table 1 are uncorrected by default, including $\lambda_{max}$ and $\Delta V$.

The dynamical age of the bubble is $0.71 \times 10^5$ yr, lower than the minimum dynamical timescale of the 37 bubbles in Taurus (Li et al. 2015), $\sim 1 \times 10^5$ yr. The dynamical age of the blue lobe of the outflow is between $0.18 \times 10^5$ yr and $0.23 \times 10^5$ yr, which is lower than the minimum dynamical timescale of the Taurus outflow survey (Li et al. 2015), $\sim 0.3 \times 10^5$ yr. For the red lobe of the outflow, the dynamical timescale is between $1.0 \times 10^5$ yr and $1.4 \times 10^5$ yr, which is close to the median value in Taurus, $\sim 1.5 \times 10^5$ yr. This is reasonable as compared to the estimated dynamical age of the Taurus molecular cloud, $\sim 10^7$ yr (Vázquez-Semadeni et al. 2018), and the dynamical ages of Taurus stellar feedback, $\sim 0.3 \times 10^4$ to $10^6$ yr (Li et al. 2015).

The two lobes of the outflow have asymmetrical dimensions and inconsistent intensities. The morphology, spatial scale, and estimated dynamical timescale of the blue lobe are significantly different from those of the red lobe of the outflow. The blue lobe has a clearer spatial distribution in the integrated intensity map in Figure 7. The red lobe is more extended. In the P–V diagram, the high-velocity gas of the blue lobe deviates from the system velocity more than that of the red lobe. Here we only discuss the formation history between the clearer blue lobe and the bubble.

### 4. Discussion

#### 4.1. Formation History

The spatial distribution of the bubble and the outflow determines their past relationship. It requires us to find the driving source of the bubble–outflow structure. At the center of the bubble, there are no identified YSOs.[11] To find the driving source, we discuss two possibilities: (1) There may be a $^{13}$CO

---
[11] We refer to the latest Spitzer YSO survey (Rebull et al. 2010).

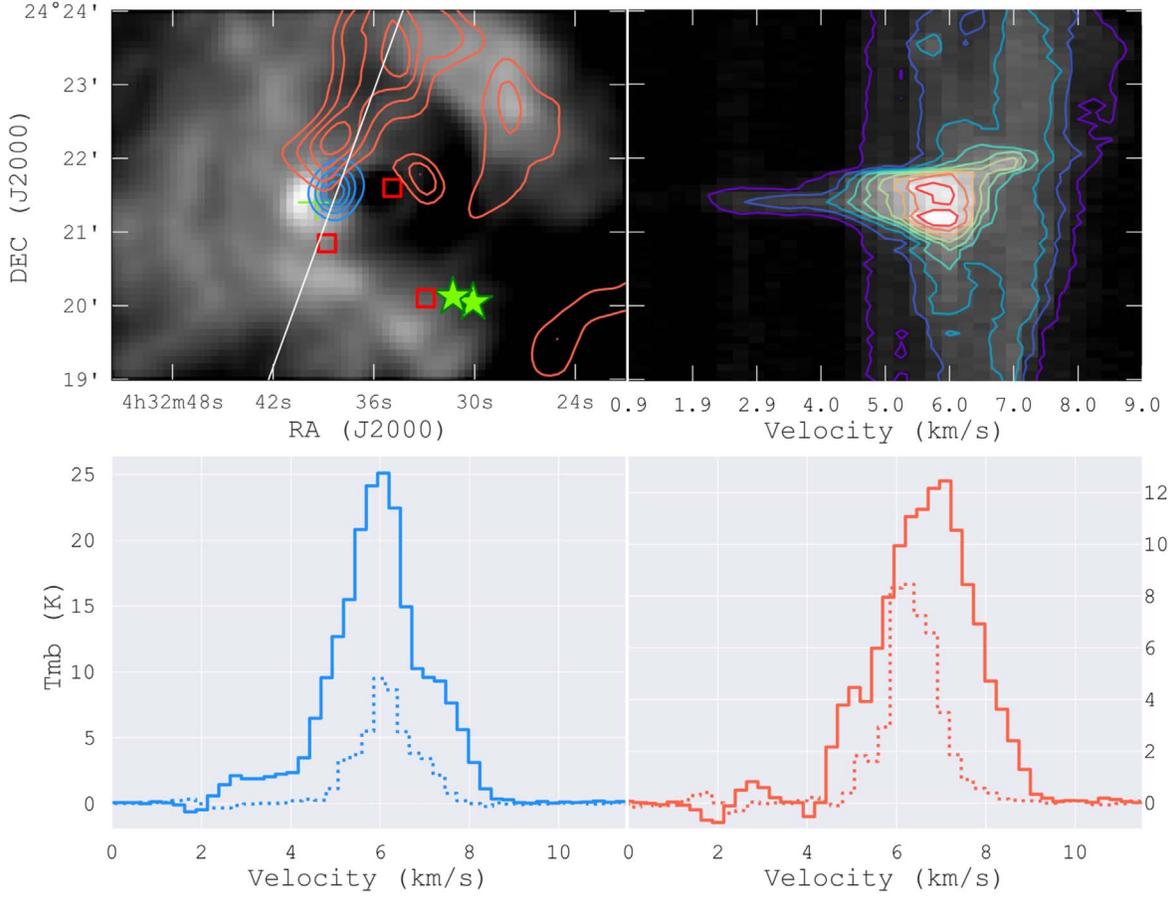

**Figure 7.** Spatial distribution, P–V diagram, and average spectra of the red and blue lobes of the outflow. Upper left panel: gray-scale image showing the integrated intensity map of $^{13}$CO $J = 2 \rightarrow 1$ in the velocity range of 5 to 7 km s$^{-1}$. The blue and red contours show the integrated intensity map of $^{12}$CO $J = 2 \rightarrow 1$, over 2–4.5 km s$^{-1}$ for the blue lobe and 7.5–9.0 km s$^{-1}$ for the red lobe. The contour levels are 20%, 40%, 60%, and 80% of the peak value for the blue lobe, and 60%, 70%, 80%, and 90% of the peak value for the red lobe. The white solid line represents a cut for the P–V diagram. The data have been smoothed to 20″. The green stars, green cross, and red squares represent the T Tauri binary stars, HH 319, and nearby HH objects, respectively. Upper right panel: P–V diagram of $^{12}$CO $J = 2 \rightarrow 1$ along the line shown above at a position angle of 70°. The contour levels are 10%, 20%, …, 90% of the peak value. Lower left panel: the average lines of the blue lobe. The solid and dotted lines represent $^{12}$CO and $^{13}$CO $J = 2 \rightarrow 1$. Lower right panel: the average $^{12}$CO and $^{13}$CO $J = 2 \rightarrow 1$ spectra of the red lobe.

**Table 1**
Physical Parameters of the Outflow and the Bubble

| Physical Parameters | Outflow Blue Lobe | Outflow Red Lobe | Bubble |
|---|---|---|---|
| $\Delta V$ (km s$^{-1}$) | 2.6 | 2.2 | 1.1 |
| $S$ (arcmin$^2$) | $0.′80 \times 0.′80$ | $4.′0 \times 4.′0$ | 11 |
| $\lambda_{max}$ (pc) | 0.046 | 0.23 | ... |
| $R_{bubble}$ (pc) | ... | ... | 0.077 |
| $N_{tot}(^{12}CO)$ ($10^{18}$ cm$^{-2}$) | 0.017 | 0.0060 | 1.3 |
| $M_{gas}$ ($M_\odot$) | 0.0039 | 0.035 | 5.2 |
| $P$ ($M_\odot$ km s$^{-1}$) | 0.0099 | 0.077 | 5.5 |
| $P_{corr}$ ($M_\odot$ km s$^{-1}$) | 0.012 | 0.096 | ... |
| $E$ ($10^{43}$ erg) | 0.025 | 0.17 | 5.8 |
| $E_{corr}$ ($10^{43}$ erg) | 0.039 | 0.26 | ... |
| $t_{dyn}$ ($10^5$ yr) | 0.18 | 1.0 | 0.71 |
| $t_{dyn,corr}$ ($10^5$ yr) | 0.23 | 1.4 | ... |
| $\dot{E}$ ($10^{31}$ erg s$^{-1}$) | 0.045 | 0.053 | 2.6 |
| $\dot{E}_{corr}$ ($10^{31}$ erg s$^{-1}$) | 0.054 | 0.062 | ... |

**Note.** All calculations are based on the main-beam temperature scale. We assume the excitation temperature to be 25 K (Li et al. 2015).

core in the center of the bubble, which is not visible in the C$^{18}$O integrated intensity map. A newborn protostar may be embedded in the core at the bubble's center. The previous YSO survey did not identify one. We analyze the mass, extinction, energy, gravitational energy, and kinetic energy of the core in Section 4.1.1. (2) The known YSOs have moved away out of the bubble. In Section 4.1.2, we discuss the proper motions of the adjacent YSOs and study their moving trajectories in the estimated dynamical timescales of the bubble and outflow.

### 4.1.1. Physical Conditions of the Core

In the bubble center, IRAM $^{13}$CO $J = 2 \rightarrow 1$ data show a dense gas distribution. We check the C$^{18}$O spectrum of this core. This core has C$^{18}$O emission, which is shown in panel (B) in the last row of Figure 9 (bottom) in Appendix A. Point B in Figure 9 represents the position of the outflow and the core. Though the integrated intensity map (in the range 4–8 km s$^{-1}$) of C$^{18}$O does not trace the core revealed by $^{13}$CO, the core can be faintly seen in the C$^{18}$O channel map at 6.16 km s$^{-1}$ (Figure 4). The large difference between the two isotopologues in revealing the core might be attributed to the outflow wings contributing to the $^{13}$CO emission, which also explain the larger, non-Gaussian line width of $^{13}$CO toward the core (spectrum B in Figure 9). Here we analyze the physical parameters of the $^{13}$CO core and discuss the possibility of protostar formation.

Normally cloud cores are prolate ellipsoids (Myers et al. 1991; Li 2002; Li et al. 2013). We fit the cloud core with an ellipse. The fitting result is expressed as $a_{core} = 0.012$ pc ($2.6 \times 10^3$ au) for the semimajor axis and $b_{core} = 0.0046$ pc ($0.94 \times 10^3$ au) for the semiminor axis. It is smaller than the typical molecular core dimension ($\sim 0.1$ pc; Myers & Goodman 1988), and the Jeans length ($\lambda_J \sim 0.83$ pc) in the Taurus molecular cloud (Li et al. 2015).

We estimate the gas mass and extinction of the core. We use $^{13}$CO $J = 2 \to 1$ integrated intensity data within the velocity interval of 3.0 to 9.0 km s$^{-1}$ to estimate the column density of the core. All calculations are based on the main-beam temperature scale. The derivation of the total column density for the core is shown in Equations (B12)–(B16) in Appendix B. The column density for the rotational upper level of the transition $J = 2 \to 1$, $N_u(^{13}$CO$)$, is $2.7 \times 10^{15}$ cm$^{-2}$ for the core. We assume that $^{12}$CO is optically thick and $^{13}$CO is optically thin, and the optical depth of $^{13}$CO can be obtained from

$$\tau(^{13}\text{CO}) = -\ln\left(1 - \frac{T^{13}(\text{CO})}{T^{12}(\text{CO})}\right) \quad (6)$$

(Li et al. 2015). The calculated optical depth of the $^{13}$CO $J = 2 \to 1$ emission is 0.44. The correction factor for the column density due to the optical depth, $f_\tau$, is 1.24. The excitation temperature, $T_{ex}$, is assumed to be 25 K (Li et al. 2015). We estimate the fraction of $^{13}$CO in the upper level, $f_{up}$, to be 0.28. We assume the abundance ratio of $^{12}$CO to $^{13}$CO is 65 (Langer & Penzias 1993). The total column density of the core can be estimated by

$$N_{tot}^{12}(\text{CO}) = 65 \frac{f_\tau}{f_{up}} N_u^{13}(\text{CO}). \quad (7)$$

The total column density of the core is $7.6 \times 10^{17}$ cm$^{-2}$. The visual extinction can be calculated from

$$A_v = \frac{N(\text{H}_2)}{9.4 \times 10^{20}} = \frac{N_{tot}(^{12}\text{CO})[\text{H}_2/\text{CO}]}{9.4 \times 10^{20}}, \quad (8)$$

where [H$_2$/CO] $= 10^4$ (Pineda et al. 2010). Based on our estimate, the maximum visual extinction of the core is 8.1 mag. This result is lower than the 16 mag extinction of the nearby core (core 11 in Pineda et al. 2010). Based on the ellipsoid shape of the core, we write the mass of the core as

$$M_{core} = \pi a_{core} b_{core} N_{tot}(^{12}\text{CO})[\text{H}_2/\text{CO}] \mu_g m(\text{H}). \quad (9)$$

The mass of the core is 0.029 $M_\odot$.

We estimate the kinetic energy and the gravitational energy, referring to previous discussion of the virial theorem (e.g., McKee & Ostriker 2007; Li et al. 2013). For an axisymmetric ellipsoid core with concentric density profiles, the gravitational binding energy can be expressed as

$$\mathcal{G} = -\frac{3}{5}\alpha\beta\frac{GM^2}{R}, \quad (10)$$

where $\alpha = (1 - a/3)/(1 - 2a/5)$ for a power-law density profile $\rho \propto r^{-a}$. We assume $a = 1.6$, as in an appropriate isothermal cloud in equilibrium (Bonnor 1956). The geometry factor $\beta = \arcsin e/e$ is determined by eccentricity $e = \sqrt{(1 - f^2)}$. The intrinsic axis ratio, $f$, should be less than the observed $f_{obs}$ under the projection of the sky plane. For our core, $f_{obs}$ is 0.37. We choose the prolate ellipsoid (Fall & Frenk 1983) and find

$$f = \frac{2}{\pi} f_{obs} \mathcal{F}_1(0.5, 0.5, -0.5, 1.5, 1, 1 - f_{obs}^2), \quad (11)$$

where $\mathcal{F}_1$ is the Appell hypergeometric function of the first kind (Li et al. 2013). We estimate that $f = 0.26$ and $\beta = 1.4$. The radius $R$ in Equation (10) is the geometric average of the semimajor and semiminor axes of the ellipse, 0.0075 pc. The gravitational constant $G$ is $6.67 \times 10^{-11}$ N m$^2$ kg$^{-2}$. We estimate the gravitational binding energy of this core to be $-9.9 \times 10^{39}$ erg. The turbulent kinetic energy of the core is

$$\mathcal{E} = \frac{3}{2} M \sigma^2, \quad (12)$$

where the velocity dispersion $\sigma$ is $0.57 \times 10^3$ m s$^{-1}$ from the Gaussian fitting of the $^{13}$CO line. We estimate a value of $2.8 \times 10^{41}$ erg for this core.

The ratio of the total kinetic energy to the gravitational binding energy is defined as the virial ratio (McKee & Ostriker 2007), expressed as

$$r_{vir} = \frac{5\sigma^2 R}{GM\alpha\beta}. \quad (13)$$

For $r_{vir} = 1$, the uniform, unmagnetized gas sphere is in virial balance ($|\mathcal{G}| = 2\mathcal{E}$). We estimate the virial ratio of this core to be $r_{vir} = 57$. Although we do not have magnetic field data, it appears that this core is definitely not gravitationally bound.

The measured non–gravitationally bound condition of the $^{13}$CO core could have two possible causes. (1) The bubble and the outflow are from the core. A newborn protostar is buried deep in the core, which generated the bubble and the outflow. This process released enough kinetic energy into its surroundings to cause the core to be in a non–gravitationally bound state. (2) The outflow and the bubble do not originate from the core, which may be in the process of being destroyed or dismantled. No protostar existed in the core to drive the bubble and the outflow. Both of the above scenarios are possible based on the estimation of the physical parameters of the core.

### 4.1.2. Nearby Known YSOs

Because of the lack of direct evidence, we can only speculate that a YSO drove the expansion of the bubble and then moved out of the bubble. To explore the origin of this structure and identify the driving source, we check the up-to-date Spitzer YSO catalog (Rebull et al. 2010). There are four possible YSOs: the common names are FY Tau, FZ Tau, CFHT-5, and MHO-8. FY Tau and FZ Tau are a pair of binary stars (Kenyon & Hartmann 1995; Kraus & Hillenbrand 2009; Akeson et al. 2019). We summarize the main information of the four YSOs in Table 2. The common names, coordinates, and YSO classifications are adopted from Rebull et al. (2010). The projected distances to the bubble, $d_{bubble}$, are the distances between the YSOs and the fitted center of the bubble ($4^h32^m35\overset{s}{.}6 +24°21'46''$).

For these four nearby possible driving sources, we search for information from Gaia Early Data Release 3 (EDR3; Gaia Collaboration et al. 2016, 2020, 2021) archive data. The proper-motion velocity ($\mu_\alpha$, $\mu_\delta$) and parallax of the four stars are taken from Gaia EDR3 (Gaia Collaboration et al. 2016, 2020, 2021). We calculate the distance to the Earth ($d_{Earth}$ in Table 2) from the parallax.

Table 2
Properties of Candidate Driving Sources

| Name | FY Tau | FZ Tau | CFHT-5 | MHO-8 |
|---|---|---|---|---|
| R.A. | $4^h32^m30\overset{s}{.}5$ | $4^h32^m31\overset{s}{.}7$ | $4^h32^m50\overset{s}{.}2$ | $4^h33^m01\overset{s}{.}9$ |
| Decl. | $+24°19'57''$ | $+24°20'02''$ | $+24°22'11''$ | $+24°21'00''$ |
| Classification | Class II | Class II | Class III | Class III |
| $d_{\text{bubble}}$ (arcmin) | 2.21 | 1.97 | 3.69 | 6.64 |
| Parallax (mas) | $7.72 \pm 0.02$ | $7.74 \pm 0.03$ | $10.26 \pm 1.85$ | $7.74 \pm 0.08$ |
| $d_{\text{Earth}}$ (pc) | $129.13 \pm 0.33$ | $129.25 \pm 0.43$ | $97.43 \pm 18.16$ | $129.13 \pm 1.36$ |
| $\mu_\alpha$ (mas yr$^{-1}$) | $6.84 \pm 0.03$ | $7.31 \pm 0.03$ | $6.49 \pm 2.40$ | $7.00 \pm 0.09$ |
| $\mu_\delta$ (mas yr$^{-1}$) | $-21.55 \pm 0.02$ | $-21.40 \pm 0.03$ | $-24.58 \pm 1.99$ | $-20.39 \pm 0.07$ |
| $\mu_{\alpha,\text{bulk\_motion}}$ (mas yr$^{-1}$) | $7.55 \pm 0.07$ | $7.55 \pm 0.07$ | $11.85 \pm 0.06$ | $7.46 \pm 0.02$ |
| $\mu_{\delta,\text{bulk\_motion}}$ (mas yr$^{-1}$) | $-20.43 \pm 0.03$ | $-20.25 \pm 0.02$ | $-31.65 \pm 0.18$ | $-19.77 \pm 0.16$ |
| $\mu_{\alpha,\text{corr}}$ (mas yr$^{-1}$) | $-0.71 \pm 0.07$ | $-0.24 \pm 0.08$ | $-5.36 \pm 2.40$ | $-0.47 \pm 0.09$ |
| $\mu_{\delta,\text{corr}}$ (mas yr$^{-1}$) | $-1.11 \pm 0.04$ | $-1.15 \pm 0.03$ | $7.07 \pm 2.00$ | $-0.62 \pm 0.18$ |

We examine the nearby YSOs that have a systematic bulk motion toward the southeast direction. Luhman (2018) used Gaia DR2 to study the systematic proper motions of stars for the Taurus molecular cloud and divided Taurus into nine regions. For one of them, dark clouds L1536, L1529 (i.e., the B18 cloud), and L1524 covered a submap of $3.5° \times 4°.25$ (see Figure 6 in Luhman 2018). In Figure 6 of Luhman (2018), there is a red population and a blue population distinguished from the parallax. Based on the spatial distribution, the four possible driving sources of the bubble in the B18 cloud here all belong to the red population. The red population in Figure 6 in Luhman (2018) has a coverage of about $2° \times 4°.25$ and has a bulk motion of $\mu_\alpha = 7.2$ mas yr$^{-1}$, $\mu_\delta = -24.1$ mas yr$^{-1}$, $\Delta\mu_\alpha = -1.0$ mas yr$^{-1}$, and $\Delta\mu_\delta = 1.5$ mas yr$^{-1}$ (see Table 2 in Luhman 2018). Considering that the error of the Gaia EDR3 data for the four YSOs is smaller than that of Gaia DR2, we choose to calculate the bulk-motion velocity of the YSOs using Gaia EDR3 (Gaia Collaboration et al. 2016, 2020, 2021).

Acquiring the bulk motions for the four possible driving sources first requires defining the region in which to select the surrounding stars. We adopt the division of this region from Luhman (2018) (red population in Figure 6) as the scope. We set the angular radius of $4°.25$ as a selection range for stars in the sky plane and line-of-sight direction, which ensures the full coverage of all Gaia sources in the B18 cloud. For each of the possible driving sources (R.A.$_0$, decl.$_0$, parallax$_0$), we use Gaia EDR3 to select the stars around it. In the sky projection plane, we select stars in the range

$$\text{R.A.}_0 - 4°.25 \leqslant \text{R.A.} \leqslant \text{R.A.}_0 + 4°.25,$$
$$\text{decl.}_0 - 4°.25 \leqslant \text{decl.} \leqslant \text{decl.}_0 + 4°.25. \quad (14)$$

In the line-of-sight dimension, star selection should convert $4°.25$ to parsecs and calculate the interval of the line-of-sight distance. This distance interval is converted into a parallax interval. Only stars with parallax in the range

$$\frac{\text{parallax}_0}{1 + 4.25\pi/180} \leqslant \text{parallax} \leqslant \frac{\text{parallax}_0}{1 - 4.25\pi/180}, \quad (15)$$

which is

$$0.93\,\text{parallax}_0 \leqslant \text{parallax} \leqslant 1.08\,\text{parallax}_0, \quad (16)$$

should be selected as the surrounding stars for the possible driving source (R.A.$_0$, decl.$_0$, parallax$_0$).

The Gaia data indicate that YSOs have systematic motions relative to other stars in the sky projection plane. The four possible driving sources of the bubble are YSOs. Thus, the four possible driving sources have systematic motions relative to stars other than YSOs. To deduce the relative motions of the four possible sources relative to the cloud, we take the median proper motion of stars excluding all YSOs as a reference. We use the following steps for the calculation. (1) For the four possible driving sources, we select all stars around them within $4°.25$ in Gaia EDR3. (2) All YSOs are eliminated from the selected stars. The catalog of YSOs is from the latest Spitzer YSO survey (Rebull et al. 2010). (3) We calculate the median proper motion of all selected stars excluding YSOs to be $\mu_{\alpha,\text{bulk\_motion}}$, $\mu_{\delta,\text{bulk\_motion}}$ (Table 2). (4) We calculate the corrected proper motion of the driving sources relative to $\mu_{\alpha,\text{bulk\_motion}}$, $\mu_{\delta,\text{bulk\_motion}}$ to be $\mu_{\alpha,\text{corr}}$, $\mu_{\delta,\text{corr}}$ (Table 2). (5) When we select stars in Gaia EDR3, we constrain their parallaxes using Equation (16). Among the four possible driving sources, CFHT-5 is 97 pc away from us, and the other three are 129 pc away from us. The selected stars around them and their bulk motions are different. We conduct the above process four times to obtain the corrected proper motions.

With the dynamical timescale of the bubble and the corrected bulk-motion velocities, we can obtain the motion trajectory, shown in Figure 8. We explore using a selection range that is larger or smaller than $4°.25$. The proper-motion trajectories of FY Tau and FZ Tau are shown as movement from the north to the south. Compared to the other two YSOs, the T Tauri binary stars, FY Tau and FZ Tau, are the most probable ones to have originated from the bubble center to the present observed position. They are more likely to explain the bubble–outflow formation than the other two YSOs in many attempts. The distance of CFHT-5 in the line-of-sight direction is 97 pc. There are relatively fewer stars around CFHT-5. Regardless of whether the selection range is greater or less than $4°.25$, CFHT-5 has a different bulk motion compared with the other three possible sources. In the case of MHO-8, many attempts have shown that it is difficult to get a trajectory from the interior of the bubble to where it is now.

T Tauri stars drive wide-angle winds, which in some cases coexist with a collimated jetlike wind component (Arce et al. 2011). They cannot move too far from their birthplace during their lifetime (about $10^6$ yr; Gomez et al. 1993). Li et al. (2015) and Magakian et al. (2002) identified the pair of T Tauri stars as the driving source of this bipolar outflow and HH 319. In conclusion, there is a possibility the T Tauri binary stars, FY Tau and FZ Tau, are the driving source of the bubble–outflow structure.

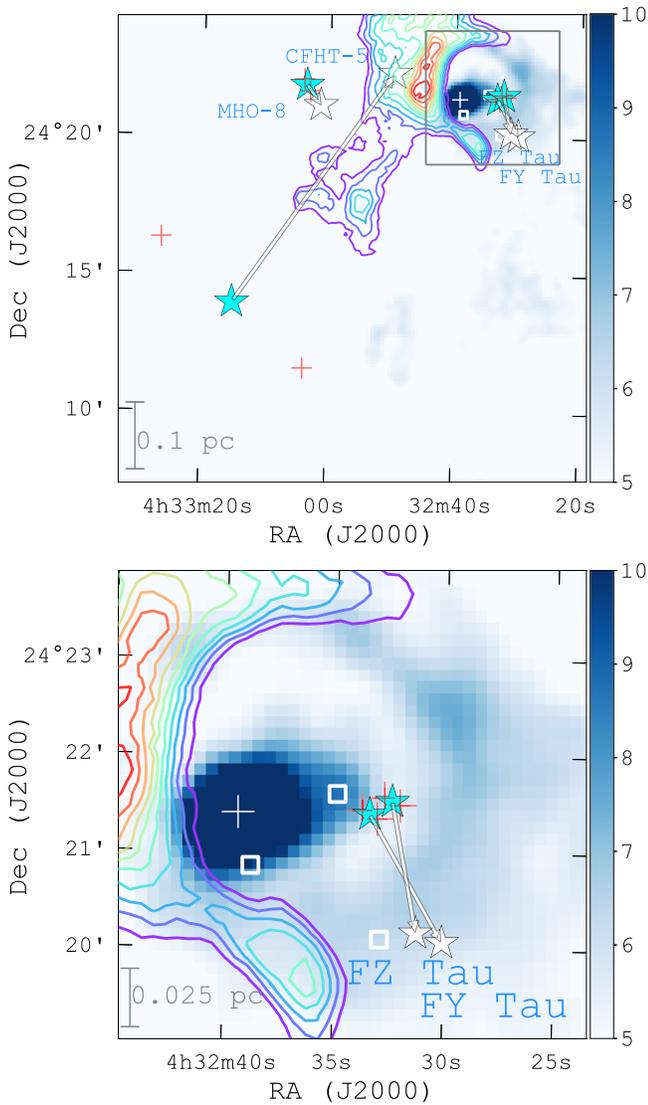

**Figure 8.** Upper panel: the proper-motion paths of the nearby YSOs in $7.1 \times 10^4$ yr (from west to east, FY Tau, FZ Tau, CFHT-5, and MHO-8). The color background is the integrated intensity map of $^{12}$CO $J = 3 \rightarrow 2$ from 5 km s$^{-1}$ to 6 km s$^{-1}$ (Y. Duan et al. 2023, in preparation). The contours are the Herschel SPIRE 250 $\mu$m data. The contour levels are 50%, 55%, ..., 95% of the peak value. The white and cyan stars represent the positions of the YSOs at present and $7.1 \times 10^4$ yr ago, respectively. The arrows show the motion paths of the YSOs. The red crosses represent the upper limit and lower limit of the corrected proper motions for the YSOs. The white cross and squares are HH 319 and other HH objects. The gray square represents the zoomed-in region shown in the lower panel. Lower panel: a zoomed-in view of the above panel, showing the proper-motion paths of the T Tauri binary (FY Tau and FZ Tau).

We have developed the following possible explanation. The T Tauri binaries (FY Tau and FZ Tau) formed this bubble and outflow. The formation of the bubble preceded the formation of the outflow. Otherwise, the process of the bubble expanding would destroy the structure of the outflow. We reconstruct the trajectories of their movements over the dynamical ages, shown in Figure 8.

Observations have demonstrated that a subset of HH outflows and jets exhibit C-symmetric bending when the source is moving through the medium (Bally & Reipurth 2001; Bally 2016). In our concept, FY Tau and FZ Tau moved from the bubble interior to their present positions. C-symmetric bending is not present in any optical images (e.g., the H$\alpha$ image in Figure 1). We consider two possible reasons for its invisibility. (1) The intensity of the outflow is changing. It may be visible initially and become weak or invisible with the passage of time. The HH object only traces gas that passed through a shock recently. It would become invisible after the post-shock gas cools (Bally & Reipurth 2001; Bally et al. 2012). (2) HH flows and objects are formed by the interaction of the YSOs with the surrounding medium. It is possible that the surrounding gas is not dense enough to produce clearly visible structures. If future data can reveal this C-symmetric bending structure, the associated source could be demonstrated to be the driving source of the bubble–outflow structure. At present, FY Tau and FZ Tau are the most probable sources to generate this bubble and outflow.

### 4.2. Comparison with Observations and Models

#### 4.2.1. The Outflow

Both bubble and outflow originate from a period of mass loss during star formation. Generally, the collimation of the stellar wind decreases with the star's evolution (e.g., Arce & Sargent 2006; Seale & Looney 2008; Velusamy et al. 2014). Recently, several models of stellar winds have been developed to explain the different types of morphology, velocity, collimation, and momentum injection (Cabrit et al. 1997). Arce et al. (2007) summarized four types of outflow models that are widely accepted: (1) wind-driven shells (e.g., Li & Shu 1996; Lee et al. 2001), (2) jet-driven bow shocks (e.g., Chernin & Masson 1995; Cliffe et al. 1996; Hatchell et al. 1998; Lee et al. 2001), (3) jet-driven turbulent flows (e.g., Canto & Raga 1991; Chernin & Masson 1995; Bence et al. 1996), and (4) circulation flows (e.g., Fiege & Henriksen 1996; Lery et al. 1999). These models are used to explain the observed spatial structures and velocity fields of outflows. Some outflows are more complex and cannot be explained by a single model. For example, a wide-angle wind and a collimating episodic wind could cooperate to produce the observed red lobe of the HH 46/47 outflow (Arce et al. 2013). Lee et al. (2002) observed some CO outflows with multipolar structures and proposed that a model combining both jet-driven and wind-driven components may best match the observation. There are some other observations. T Tauri stars traced by forbidden emission line profiles show two-velocity components: a high-velocity component that is argued to arise from a jet and a low-velocity component that might result from a disk wind (Kwan & Tademaru 1995; Li & Shu 1996).

The present outflow was identified as a bipolar outflow in a previous study (Li et al. 2015). The two lobes of the outflow traced by $^{12}$CO have different morphologies. This is reasonable and common, similar to the case of the HH 46/47 outflow (Arce et al. 2013) and the GK/GI Tau outflow (Arce & Sargent 2006). The blue lobe of the present outflow is similar to the morphology described by the jet-driven bow shock model (e.g., Chernin & Masson 1995; Cliffe et al. 1996; Hatchell et al. 1998; Lee et al. 2001; Arce et al. 2007). Limited by the low velocity resolution, we are unable to make any more determinations for the blue lobe and the red lobe of this outflow.

#### 4.2.2. The Bubble

Molecular bubbles originate from spherical or very wide angle winds and are powered by young stars in molecular clouds. Norman & Silk (1980) built a dynamical model of molecular clouds that is dominated by collisions between bubbles from the interaction of T Tauri stars with the surrounding quiescent gas. In subsequent systematic surveys of molecular bubbles in Perseus, Taurus, and Orion A giant molecular clouds (Arce et al. 2011; Li et al. 2015; Feddersen et al. 2018), powerful spherical winds from bubbles have been proved as one of the main energy origins of molecular clouds for maintaining turbulence.

In the recent bubble surveys for the three molecular clouds (Arce et al. 2011; Li et al. 2015; Feddersen et al. 2018), we only find one example of a bubble coexisting with an outflow and HH flows. In the Orion A giant molecular cloud, Feddersen et al. (2018) identified two nested expanding shells, Shell 39 and Shell 40. Shell 40 is nested inside Shell 39. A multiple stellar system, V380 Ori, was proven to drive the two molecular shells and an identified molecule outflow (Morgan et al. 1991; Feddersen et al. 2018). V380 Ori (IRAS 05339-0644) is a Herbig Ae/Be intermediate-mass emission line star, one of the youngest Herbig Ae/Be stars (Herbig 1960; Allen & Davis 2008). In contrast to our molecular bubble in Taurus, there is also a far-infrared dark cavity excavated by protostellar jets from the V380 Ori system and located 0.1 pc away from molecular Shell 40 (Stanke et al. 2010). The group associated with V380 Ori drives multiple HH flows. It can drive high-velocity large-scale HH flows, like the 5.3 pc long HH 222/1041 flow (Reipurth et al. 2013). Moreover, it can be the driving source for small-scale HH flows, like HH 1031/130 and HH 35, which may represent more recent dynamical interactions. Both types of flows play a role in shaping the shells (Feddersen et al. 2018). An accretion-driven outburst of V380 Ori could explain the observed large-scale HH flows, increased mass-loss rates, and spherical winds (Feddersen et al. 2018). In our case, there is no sign of such an outburst for the T Tauri binary stars, FY Tau and FZ Tau.

Herbig Ae/Be stars and T Tauri stars have different masses and internal structures. Unlike the low mass ($\leqslant 1.5\ M_\odot$) of T Tauri stars, the masses of Herbig Ae/Be stars are in the range of 1.5–10 $M_\odot$ (Pinzón et al. 2021). Herbig Ae/Be stars are mainly radiative and T Tauri stars are mostly fully convective (Iben 1965). However, similar observational features are found around both Herbig Ae/Be stars and T Tauri stars, such as coexisting outflows, bubbles, HH flows, and far-infrared dark cavities. FY Tau, FZ Tau, and V380 Ori demonstrate that low- and intermediate-mass stars can significantly impact the interstellar medium's physical and chemical properties.

As the possible driving source, FY Tau and FZ Tau are likely to be the first T Tauri binary found to drive a molecular outflow and a molecular bubble together. Our observation of this structure again demonstrates the potential effect of bubbles and outflows for dark clouds. Higher-resolution observations will be required to ascertain the origin of this bubble–outflow structure.

## 5. Conclusions

In this paper, we have studied a bubble–outflow structure using observations of multiple transitions of CO together with existing infrared data. We analyze the energy injection, dynamical timescale, possible driving source, and formation history of the bubble and the outflow, and existing observations and models. We draw the following conclusions:

1. We have discovered a new molecular bubble. A known bipolar outflow (Li et al. 2015) is located at the interior of the bubble. The blue lobe of the outflow is coincident with the bubble center. HH 319 is in the blue lobe.

2. The molecular bubble has a mass, momentum, and kinetic energy of 5.2 $M_\odot$, 5.5 $M_\odot$ km s$^{-1}$, and $5.8 \times 10^{43}$ erg, respectively, within a radius of 0.077 pc. The energy injection rate is $2.6 \times 10^{31}$ erg s$^{-1}$. We estimate that the bubble formed $7.1 \times 10^4$ yr ago. This bubble has the smallest radius and the shortest kinematic timescale among the known 37 bubbles in Taurus (Li et al. 2015).

3. The bipolar outflow has a total mass of 0.039 $M_\odot$. Considering a projection angle of 37° with respect to the line-of-sight direction, the outflow has a total angular momentum between 0.087 $M_\odot$ km s$^{-1}$ and 0.11 $M_\odot$ km s$^{-1}$. The energy is between $1.9 \times 10^{42}$ erg and $3.0 \times 10^{42}$ erg. The energy injection rate is between $0.98 \times 10^{30}$ erg s$^{-1}$ and $1.2 \times 10^{30}$ erg s$^{-1}$.

4. According to our estimate, the core at the center of the bubble is non–self-gravitationally bound. Its kinetic energy to gravitational binding energy ratio, $r_{\rm vir} = 2\mathcal{E}/|\mathcal{G}|$, is 57.

5. The T Tauri binary stars located southwest of the bubble may be the driving sources. After subtracting their bulk motion, we find the T Tauri binary stars (FY Tau and FZ Tau) are the most likely ones to have moved from the bubble center to their present position. Their corrected proper-motion velocities are $\mu_{\alpha,{\rm corr}} = -0.71 \pm 0.07$ mas yr$^{-1}$, $\mu_{\delta,{\rm corr}} = -1.11 \pm 0.04$ mas yr$^{-1}$ for FY Tau and $\mu_{\alpha,{\rm corr}} = -0.24 \pm 0.08$ mas yr$^{-1}$, $\mu_{\delta,{\rm corr}} = -1.15 \pm 0.03$ mas yr$^{-1}$ for FZ Tau.

We compare the existing systematic bubble surveys (Arce et al. 2011; Li et al. 2015; Feddersen et al. 2018). This is the first example of bubble–outflow coexistence in a low-mass star formation region. In Orion A, there is an example of a co-located bubble and outflow. Our discovery demonstrates the potential of bubbles and outflows in energy injection and modification of structures within dark clouds. It demonstrates the ability of the stellar wind from T Tauri binary stars to make a dramatic impact on the surrounding environment. Further observations are needed to reveal the origin of the bubble and the outflow.

This work is supported by the National Natural Science Foundation of China (NSFC; grant Nos. 11988101, 11725313, and U1931117) and the International Partnership Program of the Chinese Academy of Sciences (grant No. 114A11KYSB 20210010). This research was carried out in part at the Jet Propulsion Laboratory, which is operated by the California Institute of Technology under a contract with the National Aeronautics and Space Administration (80NM0018D0004). We are grateful to Sheng-Li Qin for his kind advice on chemistry. This work is based on observations carried out under project No. 161-18 with the IRAM 30 m telescope. IRAM is supported by INSU/CNRS (France), MPG (Germany), and IGN (Spain). This research has made use of the NASA/IPAC Infrared Science Archive, which is funded by the National Aeronautics and Space Administration and operated by the California Institute of Technology. This work is based in part on archival data obtained with the Spitzer Space Telescope, which was operated by the Jet Propulsion Laboratory, California Institute of Technology, under a contract with NASA (80NM0018D0004). Support for this work was provided by NASA. This work has made use of data from the European Space Agency (ESA) mission Gaia processed by the Gaia Data Processing and Analysis Consortium (DPAC). Funding for the DPAC has been provided by national institutions, in particular the institutions participating in the Gaia Multilateral Agreement. Herschel is an ESA space observatory with science instruments provided by European-led principal investigator consortia and with important participation from NASA.

*Facilities:* IRAM:30m, Herschel, Spitzer, Gaia, Mayall.
*Software:* GILDAS (Gildas Team 2013, http://www.iram.fr/IRAMFR/GILDAS), Starlink (Currie et al. 2014).

## Appendix A
## Emission of the OFF Position

Due to the wrong selection of the OFF position when we performed the observation, there are $^{12}$CO and $^{13}$CO emissions from the OFF position. This resulted in a weaker antenna temperature and unphysical absorption. In order to correct this problem, we obtained an additional tracking observation of this OFF position on 2019 February 25. The sensitivities of the spectra for the OFF position are 0.022 K, 0.018 K, and 0.024 K ($T_a^*$) at 230.538000 GHz, 220.398684 GHz, and 219.560358 GHz, respectively. By combining the observed CO emission of the OFF position with the original absorption for each spectrum in the raw data, we have corrected this problem. Figure 9 shows the modified spectral lines from four pixels in the map. We processed the whole image to obtain the corrected image.

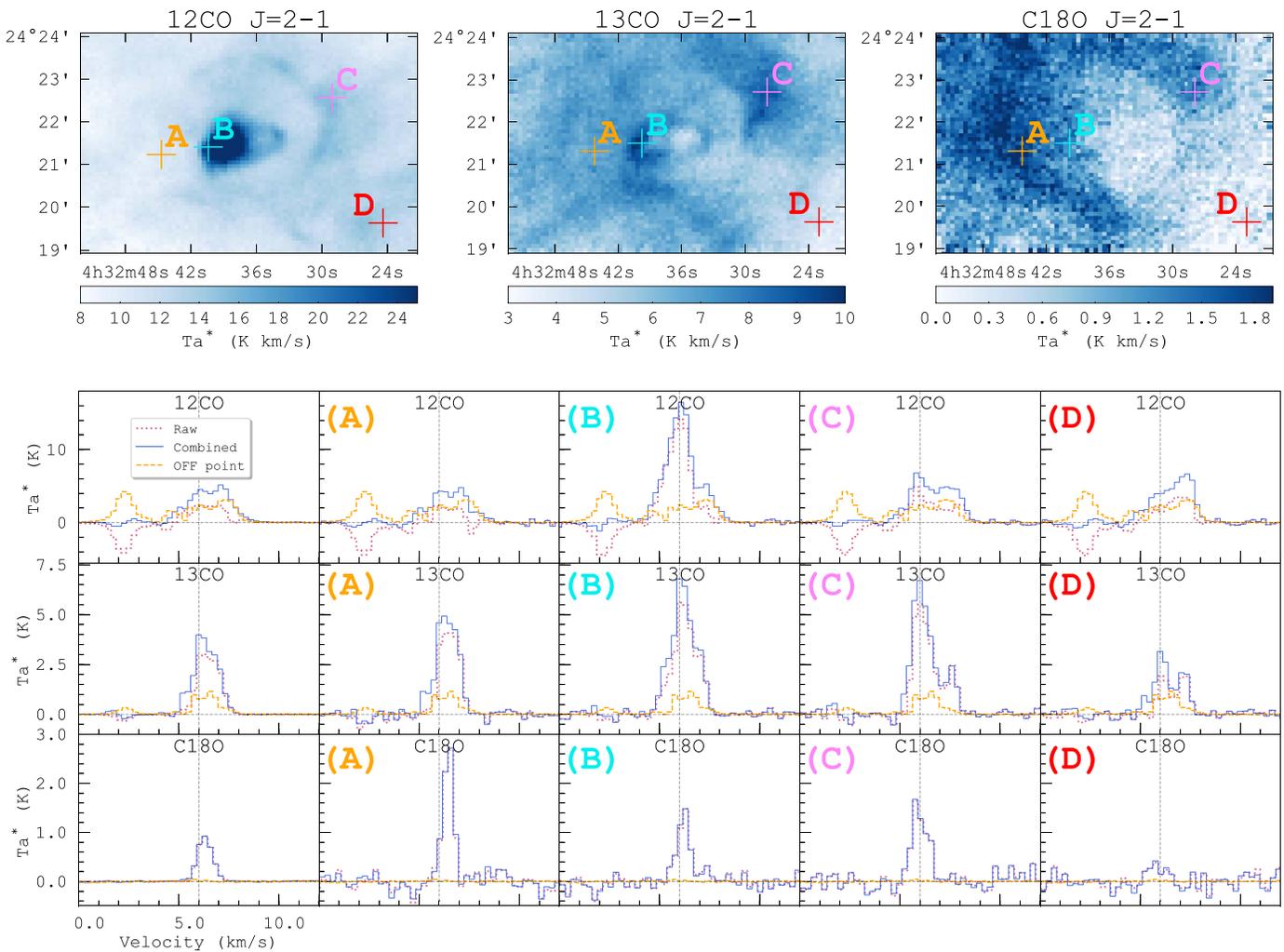

**Figure 9.** Top: integrated intensity $^{12}$CO, $^{13}$CO, and C$^{18}$O $J = 2 \to 1$ maps over 4–8 km s$^{-1}$. Four crosses mark points A, B, C, and D. Point A represents the dense gas traced by C$^{18}$O, point B the outflow and the core, point C the bubble, and point D the relatively diffuse gas. Bottom: comparison of spectral lines before and after the OFF correction at different positions. The pink dotted lines show the CO data before correction. The orange dashed line is the CO emission from the wrong OFF point. The blue solid lines are the corrected data used throughout the paper. The first column is the average spectrum of $^{12}$CO (top), $^{13}$CO (middle), and C$^{18}$O (bottom) $J = 2 \to 1$ for the total map. The velocity scale for each spectrum is the same as that for C$^{18}$O shown at lower left. Columns 2, 3, 4, and 5 represent the lines of one pixel at positions A, B, C, and D, respectively.

## Appendix B
## Physical Parameters of the Outflow and the Bubble

Referring to Li et al. (2015)'s method in the Taurus outflow–bubble survey, we adopt $^{12}$CO to estimate the physical parameters of the outflow. According to the radiative transfer equation, the intensity of the molecular spectral line should be the difference between the background temperature and the temperature of the molecular spectral line, which is

$$T_s = [T_{ex} - T_{bg}](1 - e^{-\tau_\nu}). \tag{B1}$$

Based on the definition of the optical depth and the Boltzmann equation for statistical equilibrium, the relationship between the column density of the upper energy level and the optical depth can be deduced as

$$\int \tau_\nu d\nu = \frac{8\pi^3 \nu \mu_d^2}{3hc} [e^{h\nu/kT_{ex}} - 1] N_u \tag{B2}$$

(Mangum & Shirley 2015).

Here we assume $\tau \ll 1$ and $T_{bg} \ll T_{ex}$. By substituting the optical depth expression in Equation (B2) into Equation (B1), the column density for the rotational upper level of the transition $J = 2 \to 1$ can be written as

$$N_u(^{12}CO) = \frac{8\pi k \nu^2}{hc^3 A_{ul}} \int T_s dv \tag{B3}$$

(Li 2002), where $k = 1.38 \times 10^{-16}$ erg K$^{-1}$, $h = 6.626 \times 10^{-27}$ erg s, $c = 3 \times 10^{10}$ cm s$^{-1}$, $\nu = 230.538 \times 10^9$ Hz, and $A_{ul} = 7.14 \times 10^{-7}$ s$^{-1}$ for $^{12}$CO $J = 2 \to 1$ (Rohlfs & Wilson 2004), which is the spontaneous transition rate from the upper level to the lower level.

Assuming $T_{bg} \ll T_{ex}$, we derive the optical depth directly from Equation (B1):

$$\tau(^{12}CO) = -\ln\left(1 - \frac{T_s}{T_{ex}}\right), \tag{B4}$$

where the excitation temperature, $T_{ex}$, is assumed to be 25 K from the Taurus outflow and bubble survey (Li et al. 2015). We use the main-beam temperature in Equation (B4). We can estimate the effect of the finite optical depth on the derived column density. We define a correction factor ($f_\tau$) accounting for the optical depth of $^{12}$CO,

$$f_\tau = \frac{\int \tau(^{12}CO) dv}{\int [1 - e^{-\tau^{12}(CO)}] dv} \tag{B5}$$

(Li 2002). The fraction of $^{12}$CO in the upper level is given by

$$f_{up} = \frac{g_{up} \exp(-E_{up}/kT_{ex})}{Q(T_{ex})}, \tag{B6}$$

where the statistical weight of the upper level $g_{up} = 2J_{up} + 1$. $E_{up}$ is the energy of the upper level of the observed transition, and could be used to estimate the total column density of $^{12}$CO, under the local thermodynamic equilibrium (LTE) assumption. For $^{12}$CO $J = 2 \to 1$, $E_{up} = 16.6$ K (Rohlfs & Wilson 2004). The partition function, $Q$, is defined as

$$Q(T) = \sum_{J=0}^{J=\infty} g_J \exp\left(-\frac{E_J}{kT}\right). \tag{B7}$$

The LTE partition function (for $hB/kT \gg 1$) is $Q(T) = kT/hB$, where $B$ is the rotation constant of CO, $B = 57{,}635.968$ MHz[12] (Li et al. 2015; Xie et al. 2021).

The total column density of the outflow can be estimated using

$$N_{tot}(^{12}CO) = \frac{f_\tau}{f_{up}} N_u(^{12}CO). \tag{B8}$$

We can write the gas mass as

$$M_{gas} = N_{tot}(^{12}CO)[H_2/CO]\mu_g m(H) S, \tag{B9}$$

where [H$_2$/CO] is assumed to be $10^4$, the mean molecular weight $\mu_g = 2.72$, and $m(H) = 1.67 \times 10^{-24}$ g. The momentum and energy of the outflow can be calculated from

$$P = M_{gas} \Delta V, \quad E = \frac{1}{2} M_{gas} \Delta V^2. \tag{B10}$$

The energy injection rate of the outflow and the bubble can be expressed by

$$\dot{E} = \frac{E}{t_{dyn}}. \tag{B11}$$

In the calculation of the physical parameters of the bubble with $^{13}$CO $J = 2 \to 1$ emission, the column density for the rotational upper level can be written as

$$N_u(^{13}CO) = \frac{8\pi k \nu^2}{hc^3 A_{ul}} \int T_s dv, \tag{B12}$$

where $\nu = 220.399 \times 10^9$ Hz, and $A_{ul} = 6.2 \times 10^{-7}$ s$^{-1}$ for $^{13}$CO $J = 2 \to 1$ (Rohlfs & Wilson 2004). The correction factor of the column density caused by optical depth is

$$f_\tau = \frac{\int \tau(^{13}CO) dv}{\int [1 - e^{-\tau^{13}(CO)}] dv}, \tag{B13}$$

where $\tau(^{13}CO)$ is the optical depth of the $^{13}$CO $J = 2 \to 1$ emission. If the excitation temperature is equal for $^{12}$CO and $^{13}$CO, we can obtain

$$\frac{T(^{12}CO)}{T(^{13}CO)} = \frac{1 - e^{-\tau(^{12}CO)}}{1 - e^{-\tau(^{13}CO)}}, \tag{B14}$$

where $T(^{12}CO)$ and $T(^{13}CO)$ are the brightness temperatures of $^{12}$CO and $^{13}$CO, respectively. Assuming that $^{12}$CO is optically thick and $^{13}$CO is optically thin, the optical depth of $^{13}$CO can be obtained from

$$\tau(^{13}CO) = -\ln\left(1 - \frac{T^{13}(CO)}{T^{12}(CO)}\right) \tag{B15}$$

(Li et al. 2015). The fraction of $^{13}$CO in the upper level is given by Equation (B6). For $^{13}$CO $J = 2 \to 1$, $E_{up} = 15.9$ K (Rohlfs & Wilson 2004). The rotation constant of $^{13}$CO is $B = 55{,}101.012$ MHz.[13] We assume the abundance ratio of $^{12}$CO to $^{13}$CO to be 65 (Langer & Penzias 1993). The total

---

[12] https://cdms.astro.uni-koeln.de/cgi-bin/cdmsinfo?file=e028503.cat
[13] https://cdms.astro.uni-koeln.de/cgi-bin/cdmsinfo?file=e029501.cat

column density of the bubble can be estimated as

$$N_{\rm tot}(^{12}{\rm CO}) = 65 \frac{f_\tau}{f_{\rm up}} N_{\rm tt}(^{13}{\rm CO}). \quad (B16)$$

The bubble's gas mass ($M_{\rm gas}$), momentum ($P$), energy ($E$), and energy injection rate ($\dot{E}$) are calculated by the same equations for the outflow.

## ORCID iDs

Yan Duan ⓘ https://orcid.org/0000-0003-3758-7426
Di Li ⓘ https://orcid.org/0000-0003-3010-7661
Paul F. Goldsmith ⓘ https://orcid.org/0000-0002-6622-8396
Laurent Pagani ⓘ https://orcid.org/0000-0002-3319-1021
Tao-Chung Ching ⓘ https://orcid.org/0000-0001-8516-2532
Shu Liu ⓘ https://orcid.org/0000-0001-6016-5550
Jinjin Xie ⓘ https://orcid.org/0000-0002-2738-146X
Chen Wang ⓘ https://orcid.org/0000-0001-8923-7757